\documentclass[12pt,preprint]{aastex}

\usepackage{emulateapj5}

\shorttitle{3D Simulations of Relativistic Precessing Jets}
\shortauthors{Aloy et al.}

\begin{document}

\title{3D Simulations of Relativistic Precessing Jets Probing the
Structure of Superluminal Sources}

\author{Miguel \'Angel Aloy\altaffilmark{1},
Jos\'e Mar\'{\i}a Mart\'{\i}\altaffilmark{2},
Jos\'e Lu\'{\i}s G\'omez\altaffilmark{3,4},
Iv\'an Agudo\altaffilmark{3},
Ewald M\"uller\altaffilmark{1},
Jos\'e Mar\'{\i}a Ib\'a\~nez\altaffilmark{2}}

\altaffiltext{1}{Max-Planck-Institut f\"ur Astrophysik, Karl-Schwarzschild-Str.
1, D-85741 Garching, Germany. maa@mpa-garching.mpg.de}

\altaffiltext{2}{Departamento de Astronom\'{\i}a y Astrof\'{\i}sica,
Universidad de Valencia, 46100 Burjassot (Valencia), Spain.
jose-maria.marti@uv.es}

\altaffiltext{3}{Instituto de Astrof\'{\i}sica de Andaluc\'{\i}a, CSIC,
Apartado 3004, 18080 Granada, Spain. jlgomez@iaa.es}

\altaffiltext{4}{Institut d'Estudis Espacials de Catalunya/CSIC, Edifici Nexus, c/ Gran Capit\`a, 2-4, E-08034 Barcelona, Spain}

\begin{abstract}

We present the results of a three-dimensional, relativistic, hydrodynamic
simulation of a precessing jet into which a compact blob of matter is
injected. A comparison of synthetic radio maps computed from the hydrodynamic
model, taking into account the appropriate light travel time delays, with
those obtained from observations of actual superluminal sources shows that the
variability of the jet emission is the result of a complex combination of
phase motions, viewing angle selection effects, and non-linear interactions
between perturbations and the underlying jet and/or the external medium. These
results question the hydrodynamic properties inferred from observed apparent
motions and radio structures, and reveal that shock-in-jet models may be
overly simplistic.

\end{abstract}


\section{Introduction}
Numerical hydrodynamic simulations were initially used to study
radiosources from their largest scales \citep{Bu91} to the collimation
and formation of their associated jets \citep{Ko02}. Due to the
relativistic nature of these sources, {\it relativistic} hydrodynamic
simulations have become necessary to study the superluminal sources
present in the nuclei of active galaxies and in the microquasars of
our galaxy. Improving on previous idealized analytical calculations
\citep{Al85}, these numerical methods are capable of studying the
time--dependent non--linear relativistic fluid dynamics, like for
instance the formation and propagation of shocks
\citep{Ch94,DU94,Ko96,FK96,MA99a}. The computation of the non-thermal
emission from such hydrodynamic models provides the means for a direct
comparison between observations and theory, and is hence a significant
step forward towards an understanding of the emission and structural
variability of relativistic jets
\citep{JL95,JL97,KF97,MI97,MA99b,MA00,Iv01}.

  Motivated by recent observations \citep[e.g.,][]{JL00,Ti01,We01}
revealing complex jet structure and dynamics suggestive of being the
result of multi--dimensional non--linear jet instabilities, we present
a three--dimensional relativistic hydrodynamic and emission
simulation of a precessing jet.
 
\section{Hydrodynamic model}

  Our hydrodynamic code GENESIS \citep{MA99a} is based on high
resolution shock capturing techniques, best suited to describe
ultra-relativistic flows with strong discontinuities. The code employs
the method of lines to achieve third order accuracy both in space
(using a Parabolic Piecewise Monotonic -PPM- inter-cell
reconstruction) and time (by means of a third order Total Variation
Diminishing -TVD- Runge-Kutta scheme). Guided by observational data we
simulate a jet composed of an ultra-relativistic plasma (adiabatic
index 4/3, and specific energy $50c^2$, $c$ being the speed of light)
in pressure equilibrium with an ambient atmosphere in which the jet
propagates. The plasma is injected with a bulk Lorentz factor of 6,
and is $10^3$ times lighter than the environment. We enforce a twofold
precession of the jet by imposing helical perturbations on the
injection velocity with amplitudes $\zeta_1$ ($\zeta_2$) = 0.035
(0.005), and periodicities $\tau_1$ ($\tau_2$) = 400$\,R_b/c$
(25$\,R_b/c$), where $R_b$ is the beam radius (the definitions of
$\zeta_{1,2}$ and $\tau_{1,2}$ are the same as in \cite{MA99b}). The
first, long wavelength precession ($\lambda_1 = 400\,R_b$) causes a
slight bending of the jet (within the computational domain of
longitudinal size 90$\,R_b$), while the second precession, with a much
shorter wavelength ($\lambda_2 = 25 R_b$) and smaller amplitude than
the first one, gives rise to a small helical modulation of the
beam. Both structural characteristics are observed in many
astrophysical sources \citep[e.g., 3C~120;][]{JL00}.

  According to the recent observations of \cite{Al02} radio components
are generated by the injection into the jet of material coming from
the inner accretion disk. Motivated by these observations, we perturb
the jet by injecting a blob of relativistic particles through the jet
inlet during 0.8$\,R_b/c$ in the frame attached to the source
(LAB-frame).  The blob, initially a cylindrical region of thickness
0.8$\,R_b$ and width 1.0$\,R_b$, is four times denser than the jet
plasma, but has the same velocity and internal energy per particle. As
the injected perturbation travels downstream it spreads asymmetrically
along the beam and splits into two regions (Fig.~\ref{fig1}) that have
distinct observational signatures (\S~\ref{emis}). The fact that
injected jet perturbations experience substantial evolution is
general, i.e., is not specific to the perturbation that we have
chosen. Indeed, a similar hydrodynamic evolution should be 
\begin{center}
\epsscale{0.90}
{\plotone{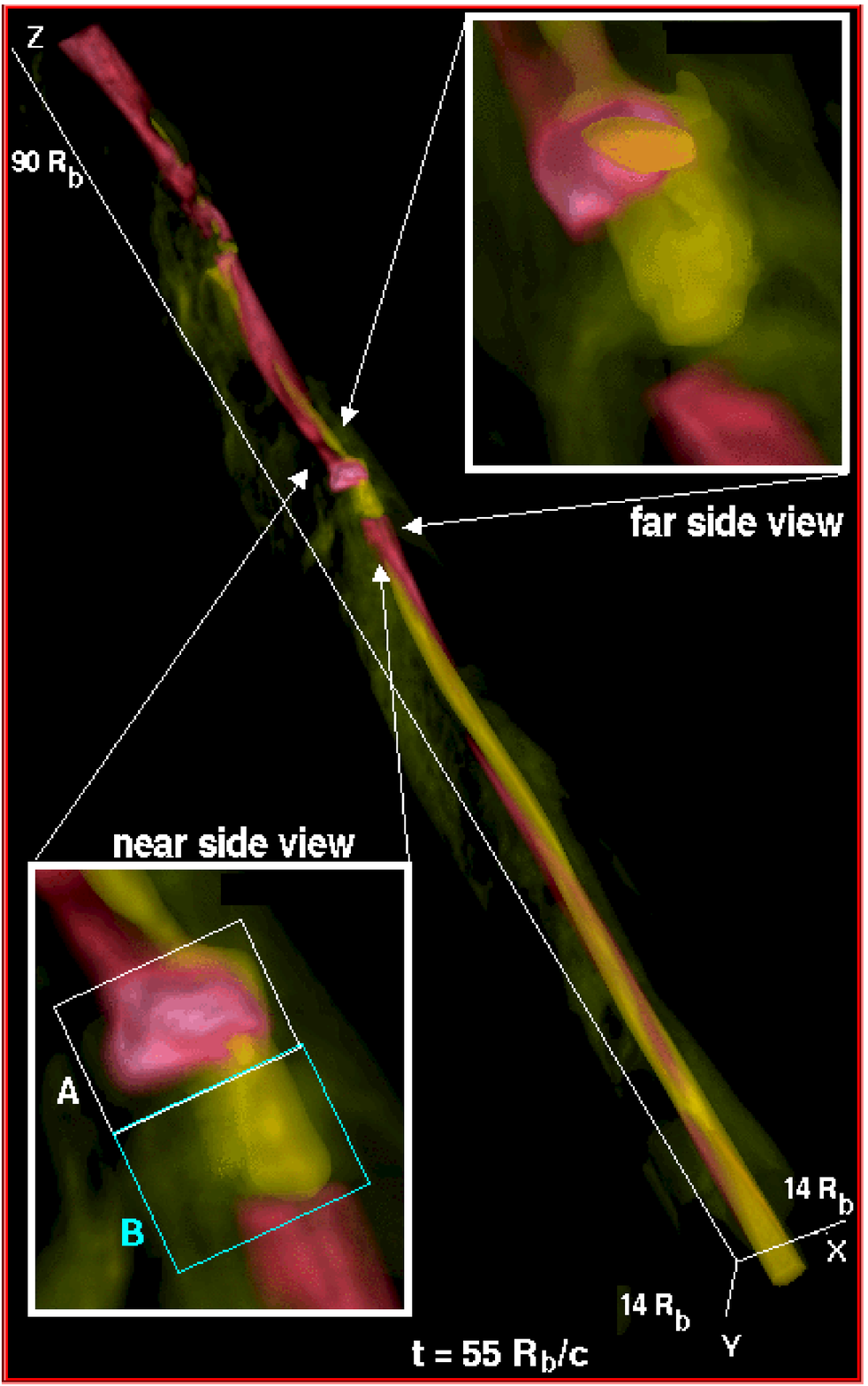}}
\figcaption{Three dimensional ray cast view of the simulated jet in
the LAB-frame. The image is produced by ray tracing the Lorentz factor
and the pressure, assigning an opacity to each volume element
proportional to the magnitude of each variable.  The computational
domain is a cuboid of physical size $14\times14\times90$ beam radii
covered by a uniform Cartesian (X,Y,Z) grid of 168x168x1080
zones. Those parts of the fluid having Lorentz factors larger than the
bulk Lorentz factor of the beam in the injection nozzle ($W_b = 6$)
show up as pink regions. Strong yellowish green colors encode the
energy density times the beam mass fraction ($X=\rho_b/\rho$, $\rho_b$
and $\rho$ being the densities of beam and of the fluid, respectively)
and diffuse greenish shades the energy density itself. In both cases,
the intensity is a measure of the corresponding quantity's value
(brighter meaning larger). The two insets are zooms of the
perturbation, the bottom one also showing regions A and B defined in
the text.\label{fig1}}
\end{center}
ex\-pec\-ted
for any perturbation that either does not match the shock jump
conditions at its boundaries or propagates along a non-uniform
beam. An idealized 1D model of the perturbation allows us to explain
the basic features of its evolution. The original state disappears
after $\approx 15\,R_b/c$ and displays a noticeable split after
30$\,R_b/c$ and beyond in the LAB-frame (Fig.~\ref{fig2}). In the 3D
simulation the evolution of the perturbation is also affected by its
interaction with the external medium and changing beam conditions
downstream.  The front region (A) of the original blob shows the
largest Lorentz factor ($\geq9$) and relatively small energy density,
while the back region (B) possesses the largest energy density but the
smallest Lorentz factor ($\approx 4.5$).

  Initially the perturbation is ballistic as a consequence of its high
relativistic density ($\approx 9.7$ times that of the external medium)
and moves in a straight trajectory until
\begin{center}
\epsscale{0.75}
\plotone{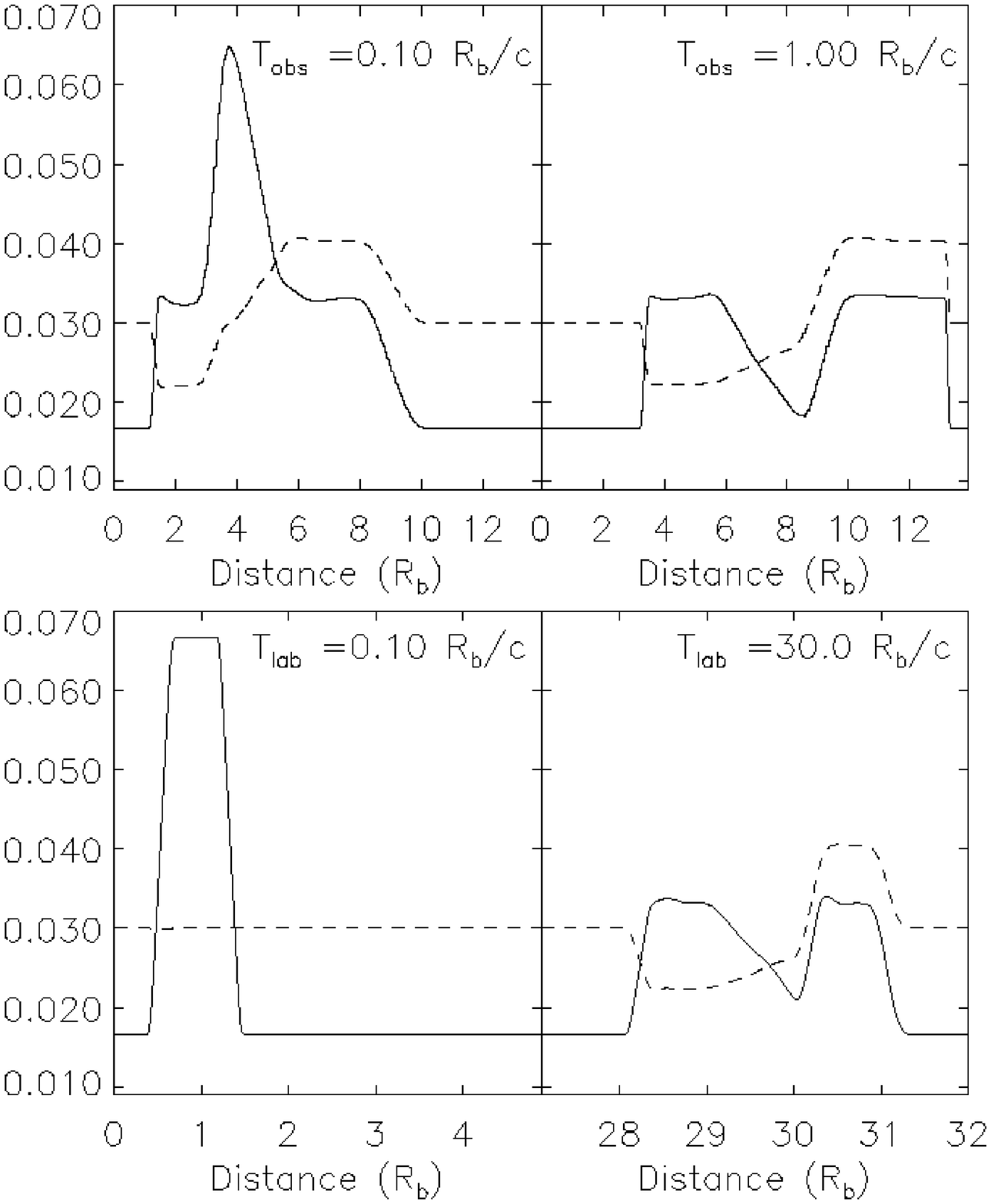}
\figcaption{Snapshots of the 1D evolution of the energy density (solid
line) and the Lorentz factor (divided by 200; dashed line) of a square
component in a uniformly moving background medium as seen in the
O-frame for a viewing angle of $15^{\circ}$ (top panels) and in the
LAB-frame (bottom panels). Physical parameters of both background and
perturbation are the same as in the 3D simulation. The quasisteady
state (rightmost panels) is very similar in both frames, but due to
the time dilation effect the apparent size is much larger in the
O-frame.
\label{fig2}}
\end{center}
it impacts on the bent walls
of the beam, between times 20 and 40$\,R_b/c$. This causes an increase
of the internal energy density and a reduction of the Lorentz factor
in the region of maximum shear (Fig.~\ref{fig3}b, \ref{fig3}c). The
heating is a consequence of the compression of the perturbation, which
is enhanced by the presence of a standing shock in the beam
(Fig.~\ref{fig3}a).  The interaction also leads to the formation of a
conical shaped bow shock that continues along with the perturbation
expanding at $\sim 0.2\,c$, which is seen in Fig.~\ref{fig3}a as a
strong increase of the energy density emerging from the component and
forming a small angle with the beam.
 In the bottom part of the component the bow shock is weaker.  The
effect of the collision against the edge of the beam is relatively
small because of the very small impact angle of $\sim 5^{\circ}$. A 2D
analytic modeling of the impact shows the generation of a rarefaction
instead of a shock for impact angles smaller than $\sim
35^{\circ}$. [This is a genuine relativistic effect
\citep{Po00,RZ02}.] After $\sim 55\,R_b/c$ the perturbation is no
longer ballistic and no longer covers the whole beam's width.

\section{Emission}
\label{emis}
 
We compute the synchrotron emission from the hydrodynamic model
assuming that the magnetic energy density is proportional to the
particle energy density and that this proportionality remains constant
throughout the whole computational domain \citep[as in][]{JL95}. We
also assume that magnetic field is dynamically negligible and we
account for the appropriate relativistic effects \citep[including
light travel time delays as in][]{JL97}, aimed to obtain synthetic
radio maps (Fig.~\ref{fig4}) that can be compared with observations of
superluminal sources. To relate emission features in the observer's
frame (O-frame;
\begin{center}
\epsscale{0.85}
\plotone{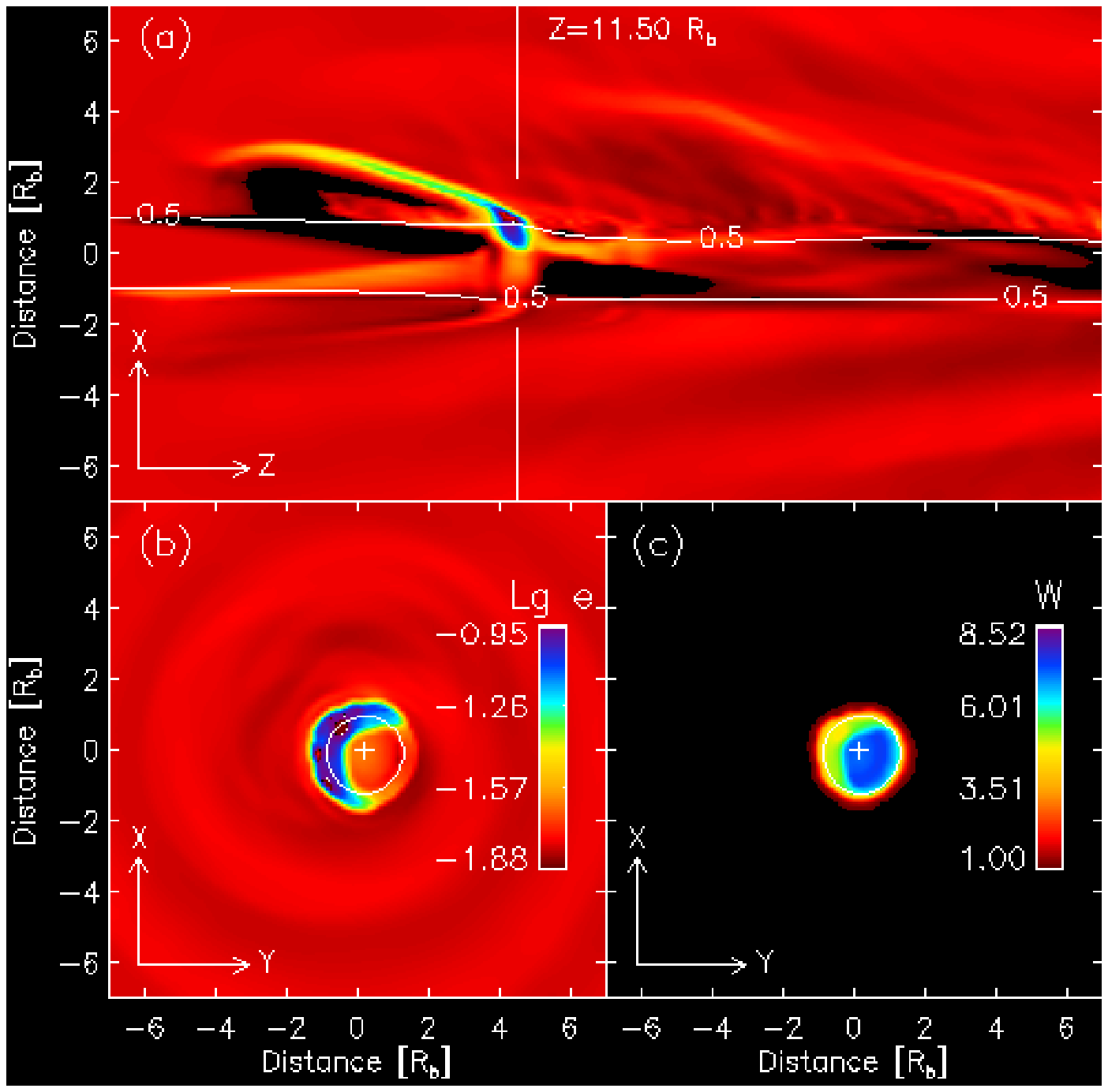}
\figcaption{Longitudinal slice through part of the jet showing the
logarithm of the energy density, e, (in arbitrary units) when the
perturbation passes through the first recollimation shock (a). The
bottom panels show the logarithm of the energy density (b; with the
same scale as the panel a) and the bulk Lorentz factor, W, (c) for
a transverse slice at z=11.5 $R_b$ (vertical white line in panel
a). In all the panels the white lines are isocontours of beam mass
fraction (see caption of Fig.~\ref{fig1}) 0.5 roughly separating the
beam from the external medium, the vertical axis provides the scale in
$R_b$, and the crosses mark the position of the jet axis if it were
axisymmetric. The spiral structure present in the ambient medium (b)
is caused by the precession of the jet.\label{fig3}}
\end{center}
 Fig.~\ref{fig4}) with their hydrodynamic counterparts, we provide a
space-time diagram (Fig.~\ref{fig5}).

Figure~\ref{fig4} shows the appearance of a large region of increased
emission, associated with the imposed perturbation, passing by several
standing knots (S1, S2, S3) caused by jet recollimation shocks. The
large extension of this new emitting region is caused by the light
travel time delays between the front and back of the perturbation,
which stretch it by a factor $(1-\beta \cos \theta)^{-1}$ ($\beta$ and
$\theta$ being the velocity --in units of $c$-- of propagation and the
viewing angle, respectively).  As a consequence, the structure of the
perturbation is magnified leading to brightness distribution
variations within the component in the O-frame.  This may have
significant implications when interpreting the observations of
superluminal sources. First of all, the quite common practice of
identifying brightness peaks in radio maps with components may be
misleading, as the component selected in this way may only be one of
many brightness features caused by a single perturbation and may not
be related to any physical structure on its own. Moreover, a component
may result from different regions of the perturbation having different
hydrodynamic properties (Fig.~\ref{fig4}). For instance, identifying
the brightness peak in the newly ejected emitting region in
Fig.~\ref{fig4} as component M leads to variations in its apparent
motion from 1.4 to 4.2$\,c$. The LAB-frame pattern Lorentz factors of
the perturbation inferred from these values are 2 to 4.3, which cannot
be associated to any flow or pattern speeds in the jet. On one hand,
they are not related to the motion of any fluid element as the bulk
Lorentz factors (Fig.~\ref{fig2}) in the perturbation dynamically vary
from the initial value of 6 (uniform within the perturbation) to
values ranging from 4 (region B) to 9 (region A). On the other hand,
the
\begin{center}
\epsscale{0.67} \plotone{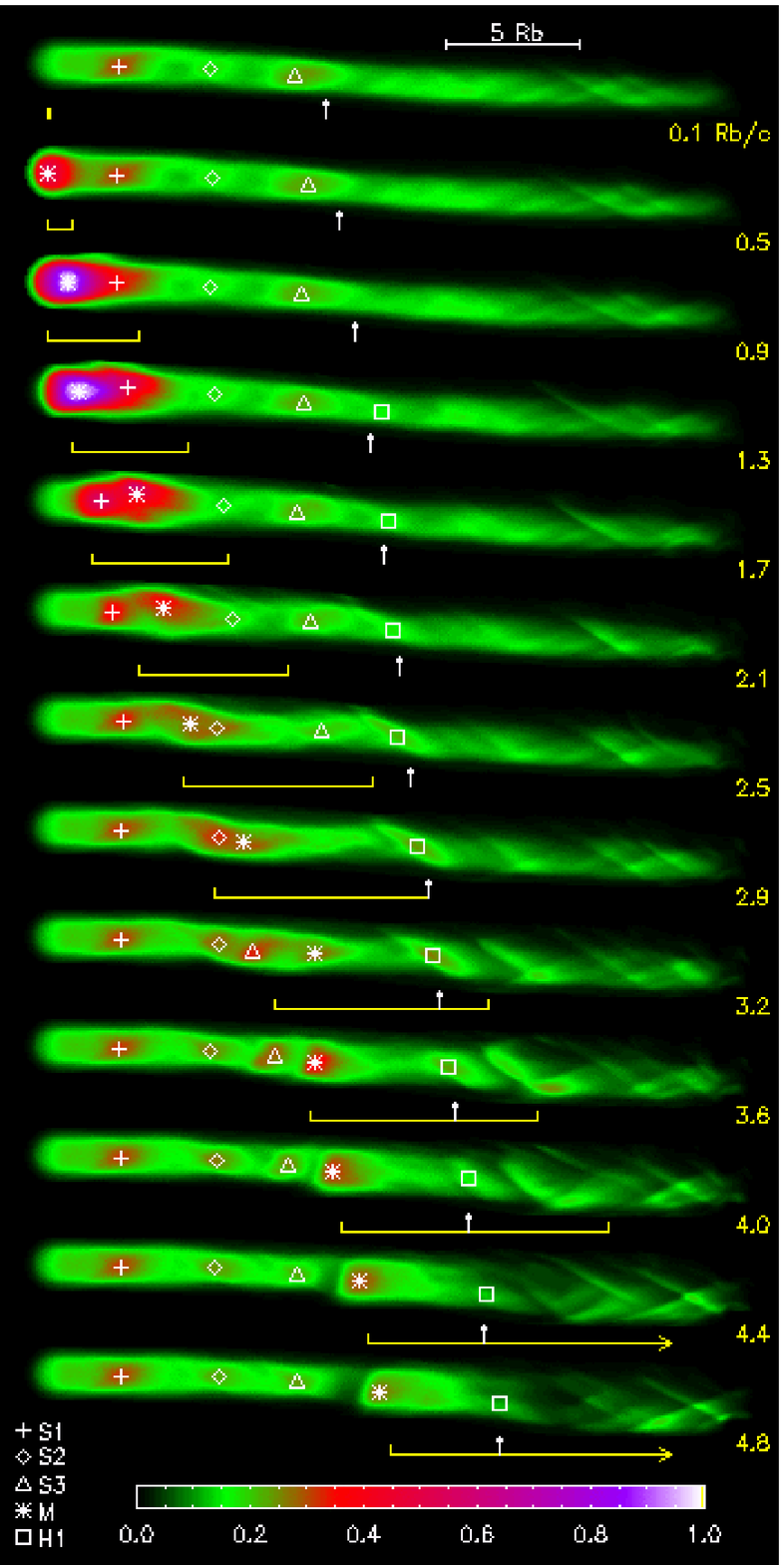}
\figcaption{Time sequence of the simulated radio maps (total intensity
in arbitrary units using a square root brightness scale) as seen in
the O-frame. The epoch is shown at the right of each snapshot. The
maps are computed for a viewing angle of $15^{\circ}$ and an optically
thin frequency of $22\,$GHz. Yellow underbrackets indicate the
extension of the imposed hydrodynamic perturbation (For the last two
epochs arrows indicate that the front of the perturbation is
undetermined.) The left and right bounds are associated to the red
triangles and asterisks in the inset of Fig.~\ref{fig5}, respectively.
Knots in the emission pattern (S1, S2, S3, M and H1) are marked with
different symbols, and a white arrow gives the position of component
H1, as determined from Fig.~\ref{fig5}.
\label{fig4}}
\end{center}
 pattern motions in the perturbation that correspond to the shocks
confining regions A and B have Lorentz factors of 7 and 4,
respectively, which differ significantly from the estimated
values. The proper motion of M is almost constant during the last four
epochs and equals 1.7$\,c$, corresponding to a LAB-frame pattern
Lorentz factor of 2.2, which is only slightly smaller than the Lorentz
factor of the back shock of region B during these epochs ($\approx
2.7$; Fig.~\ref{fig5}).

  The internal brightness distribution of the superluminal component M
also changes due to the interaction of the perturbation with the
ambient medium and the jet. The azimuthal position of the region of
maximum interaction (corresponding to the maximum energy density in
region B) varies with time because the perturbation moves (during its
ballistic epoch) along a straight trajectory while the beam is
helical. The resulting interaction, together with the squeezing of the
perturbation when it passes through standing knots in the jet, leads
to transverse motions of the component's emission peak. A similar
behavior has been observed in the jet of 3C~120 \citep{JL01},
supporting the idea of a precessing jet in this source and of
interactions between components and the external medium.

\begin{center}
\epsscale{0.92} 
\plotone{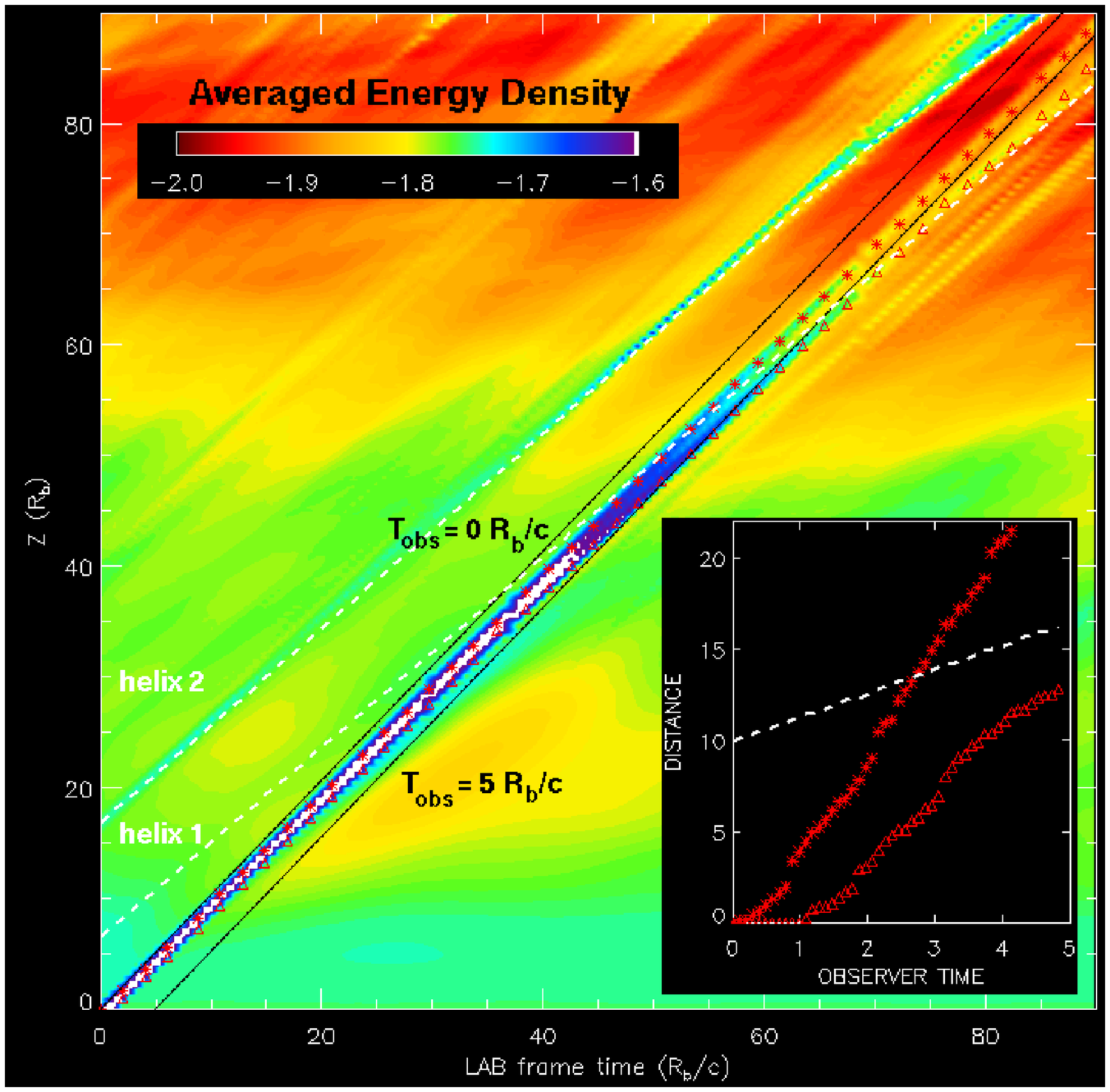} 
\figcaption{Space-time diagram showing the evolution of the energy
density along the jet. Every vertical pixel line is a snapshot of the
jet's longitudinal structure averaged across azimuthal slices. The
trajectories of two ``elbows" (i.e., local bends of the helix that have
evolved non-linearly and form kinks in the beam where the energy
density and the density are large and leave a trace in the space-time
diagram and in the radio maps of Fig.~\ref{fig4}) of the secondary
helix (helix 1 and helix 2; white dashed lines) of the helical jet
propagate with pattern Lorentz factors between 2.0-2.6. The
perturbation has average bulk Lorentz factors of 3.7 and 7.1 for the
back (B) and front (A) regions, respectively. The back shock of region B
decelerates between 40 and 55 $R_b/c$ reducing its pattern Lorentz
factor to 2.7. The area between the two black lines (isocrones in the
O-frame corresponding to times $0 R_b/c$ and $5 R_b/c$) covers the
whole time evolution of Fig.~\ref{fig4}. The propagation of the
component H1 (= helix 1) and of the main perturbation (red asterisks:
front shock of region A; red triangles: back shock of region B) are
illustrated both in the LAB- and in the O-frame (inset).\label{fig5}}
\end{center}

  While region B generates the most significant features in the radio
maps, no observable knot is produced by region A because Lorentz
factors are higher in this region and the radiation is thus beamed
into a cone smaller than the viewing angle. A larger viewing angle
would beam region A and the resulting component would present a faster
superluminal motion. Hence, the jet viewing angle may introduce a
selection effect determining which parts of the perturbation appear as
superluminal components (with different apparent speeds) in observed
radio maps.

  In contrast to M, the component H1 suddenly appears at some distance
further downstream in the jet (Fig.~\ref{fig4}). This pop-up component
({\it PUC}) results from the non-linear evolution of the secondary helical
motion of the beam, and hence does not correspond to any bulk fluid
motion. Its apparent speed is 1.4$\,c$, which agrees with its
LAB-frame pattern Lorentz factor of 2.0 (Fig.~\ref{fig5}). PUCs occur
when a relativistic perturbation (observable or not) causes enhanced
beamed emission which suddenly becomes observable. In our model the
enhanced emission is produced by shock heated matter at the edge of
region A of the perturbation, and the beaming is produced by the
relativistic flow speeds in that region. The pop-up effect arises when
the emitting region moving along the helical beam turns towards the
observer, and the edge of the beamed radiation suddenly crosses the
line of sight. PUCs appear to have been observed in actual sources
\citep[0420--014;][]{Zh00} and provide a good example of a coupling
between phase (helical) and bulk (the fluid in region A) motions
leading to the formation of new components.

\section{Conclusions}

  Our simulation shows that the non-linear hydrodynamic evolution of
perturbations can determine the observed radio emission properties of
superluminal sources, and that the interpretation of observed radio
maps is error-prone when naively associating single shocks to
superluminal components. Indeed, if the radio components of actual
sources are correctly represented by those studied in our model, most
observable features should not be related to fluid bulk motions, but
instead to a complex combination of bulk and phase motions, viewing
angle selection effects, and non-linear interactions between
perturbations and the underlying jet and/or ambient medium. Further
simulations spanning a wide range of the relevant parameters are
underway.

\begin{acknowledgements}
M. A. A. acknowledges the EU-Commission for a fellowship
(MCFI-2000-00504). This research was supported in part by the Spanish
Direcci\'on General de Enseñanza Superior grant AYA-2001-3490 and through an
agreement between the Max-Planck-Gesellschaft and the Consejo Superior de
Investigaciones Cient\'{\i}ficas.
\end{acknowledgements}


\begin{thebibliography}{}

\bibitem[Agudo et al.(2001)]{Iv01}Agudo, I., G\'omez, J. L., Mart\'{\i},
J. M., Ib\'a\~nez, J. M., Marscher, A. P., Alberdi, A., Aloy, M. A., \&
Hardee, P. E. 2001, \apjl, 549, L183

\bibitem[Aloy et al.(2000)]{MA00}Aloy, M. A., G\'omez, J. L., Ib\'a\~nez,
J. M., Mart\'{\i}, J. M., \& M\"uller, E. 2000, \apjl, 528, L85

\bibitem[Aloy et al.(1999b)]{MA99b}Aloy, M. A., Ib\'a\~nez, J. M., Mart\'{\i},
J. M., G\'omez, J. L., \& M\"uller, E. 1999b, \apjl, 523, L125

\bibitem[Aloy et al.(1999a)]{MA99a}Aloy, M. A., Ib\'a\~nez, J. M.,
Mart\'{\i}, J. M., \& M\"uller, E. 1999a, \apjs, 122, 151

\bibitem[Burns, Norman, \& Clarke (1991)]{Bu91} Burns, J. O., Norman, M. L.,
\& Clarke, D. A. 1991, Science, 253, 522

\bibitem[Duncan \& Hughes(1994)]{DU94}Duncan, G. C., \& Hughes, P. A. 1994,
\apjl, 436, L119

\bibitem[Falle \& Komissarov(1996)]{FK96}Falle, S. A. E. G., \& Komissarov,
S. S. 1996, \mnras, 278, 586

\bibitem[G\'omez et al.(2001)]{JL01}G\'omez, J. L., Marscher, A. P, Alberdi,
A., Jorstad, S. G., \& Agudo, I. 2001, \apjl, 561, L161

\bibitem[G\'omez et al.(2000)]{JL00}G\'omez, J. L., Marscher, A. P, Alberdi,
A., Jorstad, S. G., \& Garc\'{\i}a-Mir\'o, C. 2000, Science, 289, 2317

\bibitem[G\'omez et al.(1997)]{JL97}G\'omez, J. L., Mart\'{\i}, J. M.,
Marscher, A. P., Ib\'a\~nez, J. M., \& Alberdi, A. 1997, \apjl, 482, L33

\bibitem[G\'omez et al.(1995)]{JL95}G\'omez, J. L., Mart\'{\i}, J. M.,
Marscher, A. P., Ib\'a\~nez, J. M., \& Marcaide, J. M. 1995, \apjl, 449, L19

\bibitem[Koide, Nishikawa, \& Mutel(1996)]{Ko96}Koide, S., Nishikawa, K., \&
Mutel, R. L. 1996, \apjl, 463, L71

\bibitem[Koide et al. (2002)]{Ko02}Koide, S., Shibata, K.,
Kudoh, T., \& Meier, D. 2002, Science, 295, 1688 

\bibitem[Komissarov \& Falle(1997)]{KF97}Komissarov, S. S., \& Falle,
S. A. E. G. 1997, \mnras, 288, 833

\bibitem[Marscher \& Gear(1985)]{Al85}Marscher, A. P. \& Gear, W. K. 1985,
\apj, 298, 114

\bibitem[Marscher et al.(2002)]{Al02}Marscher, A. P., Jorstad, S. G., G\'omez,
J. L., Aller, M. F., Ter\"asranta, H., Lister, M. L., \& Stirling, A. M. 2002,
\nat, 417, 625

\bibitem[Mart\'{\i}, M\"uller, \& Ib\'a\~nez(1994)]{Ch94}Mart\'{\i}, J. M.,
M\"uller, E., \& Ib\'a\~nez, J. M. 1994, \aap, 281, L9

\bibitem[Mioduszewski, Hughes, \& Duncan(1997)]{MI97}Mioduszewski, A. J.,
Hughes, P. A., \& Duncan, G. C. 1997, \apj, 476, 649

\bibitem[Pons, Mart\'{\i}, \& M\"uller(2000)]{Po00}Pons, J. A., Mart\'{\i},
J. M., \& M\"uller, E. 2000, J. Fluid Mech., 422, 125

\bibitem[Rezolla \& Zanotti(2002)]{RZ02}Rezolla, L., \& Zanotti, O. 2002, PRL,
89, 114, 501

\bibitem[Tingay, Preston, \& Jauncey(2001)]{Ti01}Tingay S. J., Preston R. A.,
\& Jauncey D. L. 2001, \aj, 122, 1697

\bibitem[Wehrle et al.(2001)]{We01}Wehrle, A. E., Piner, B. G., Unwin, S. C.,
Zook, A. C., Xu, W., Marscher, A. P., Ter\"asranta, H., \& Valtaoja, E. 2001,
\apjs, 133, 297

\bibitem[Zhou et al.(2000)]{Zh00}Zhou, J. F., Hong, X. Y., Jiang, D. R., \& 
Venturi, T. 2000, \apj, 541, L13

\end{thebibliography}
\end{document}